# Study on Timing Performance of a Readout Circuit for SiPM

Liwei Wang, Yonggang Wang, Qiang Cao, Yong Xiao and Chong Liu

*Abstract*—In recent years, SiPM photoelectric devices have drawn much attention in the domain of time-of-flight-based positron emission tomography (TOF-PET). Using them to construct PET detectors with excellent coincidence time resolution (CTR) is always one of research focus. In this paper, a SiPM readout pre-amplifier based on common-base current amplifier structure followed by a Pole-Zero (PZ) compensation network is constructed, and the main factors that affect the timing performance of the PET detector are investigated. By experimental measurement, we found that the CTR is heavily related to the bandwidth of the amplifier, bias voltage of SiPM, comparator threshold, and PZ network parameter. The test setup has two detectors, one with LYSO crystal (3 mm × 3 mm × 10 mm) coupled with a Hamamatsu SiPM (S12642-0404), and the other with LaBr3 coupled to a PMT-R9800. After the optimization of the readout circuit with related factors, the CTR between the two detectors is measured as 266ps FWHM. The test result is a helpful guideline for the readout ASIC chip design in our next step.

## I. Introduction

POSITRON emission tomography (PET) is a type of nuclear medicine imaging technique that detects a pair of photons to get the position of short-lived radioisotopes. With the introduction of time-of-flight (TOF) technology, TOF-PET has been the development direction of PET system, because it can eliminate accidental coincidences and increase signal-to-noise ratio by using events arrive time [1]. Coincidence time resolution (CTR) of a pair of detectors is an important performance of PET detectors.

Silicon photomultiplier (SiPM), also known as multi-pixel photon counter (MPPC), is now replacing PMTs in TOF-PET detectors due to its features of high integration, low operation voltage and immunity to magnetic fields. However, to gain high timing performance, the intrinsic properties of SiPM brings new challenges to the readout circuit as well. Because SiPM has a big output capacitance, namely 320pf for the SiPM used in our experiment, the first stage should be a current amplification with a very low input impedance and high input bandwidth [2]. Other properties, such as photon detection efficiency (PDE), gain, and dark count rate (DCR) are all functions of bias voltage which may affect the timing performance of SiPM in some way. Higher PDE and gain can make SiPM output more charge from one event [3], which can improve CTR. But the increased of dark count is harmful to timing. During an avalanche, if the emitted photons generated by accelerated carriers can initiate other avalanche in neighboring pixels, the signal amplitude of dark count will increase by several times. More dark count and higher noise amplitude will affect signal baseline [4] and cause spurious triggering. In practical readout electronics systems, leading-edge discriminator (LED) is favored for its simplicity. However, the timing performance of LED is sensitive to baseline stability. In LED circuit, although the threshold is chosen above dark count level, the noise from dark count still decrease the CTR because the long falling edge of dark count signal makes the baseline unstable. Pole-Zero (PZ) compensation circuit has been used in SiPM readout system [4] and its effectiveness was improved. In this paper, we are studying a SiPM readout circuit with optimization of bias voltage, timing threshold and dark count shaping parameter to gain the best timing performance. The test results show that the whole readout circuit is suitable for ASIC integration in our next step.

## II. Readout Circuit and Experimental Setup

The pre-amplifier consists of a common-base amplifier and a PZ compensation network, which is shown in Fig. 1. The biasing network of bipolar junction transistor (BJT, BFS17W) consists of resistors marked as $R_{e1}$, $R_{e2}$ and $R_c$. The quiescent point did not change throughout the experiment. $R_{e2}$, C and $r_{in}$, i.e. the input resistance of common-base amplifier, constitute the PZ compensation structure. The output signal is read from the collector resistance.

Fig. 2 shows the CTR measurement setup performed for two

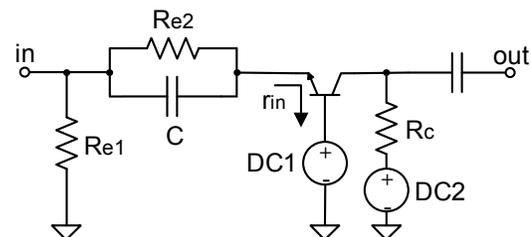

Fig. 1. Schematic diagram of the pre-amplifier based on common-base amplifier and PZ compensation.

Manuscript received February 2 2018. This work was supported in part by the National Natural Science Foundation of China (NSFC) under Grants 11475168 and 11735013.

Authors are with the Department of Modern Physics, University of Science and Technology of China, Hefei, Anhui 230026, China. Corresponding author: wangyg@ustc.edu.cn.

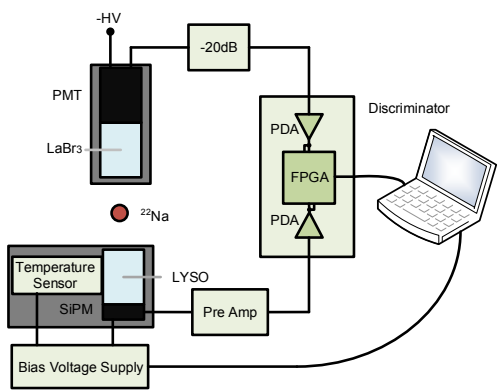

Fig. 2. Schematic representation of coincidence time measurement setup.

detect modules, one a test detector module consisting of a LYSO crystal (3 mm × 3 mm × 10 mm cuboid) optically coupled to a channel of SiPM array (S12642-0404 from Hamamatsu), and the other a coincidence detector constructed by a block of LaBr3 crystal (Ø22 mm × 10 mm cylinder) coupled to R9800 PMT. All surfaces of LYSO crystal were polished and wrapped with a layer of Tyyec paper except the surface facing to the SiPM. The programmable high-voltage power supply (C11204 from Hamamatsu) of SiPM provides temperature compensation function connecting with the temperature sensor attached near SiPM in lightproof box. The signals coming out from the SiPM were amplified by the pre-amplifier mentioned above. And the anode signals coming out from R9800 PMT were attenuated by a -20 dB attenuator for amplitude adjustment. The amplified SiPM signals and attenuated PMT signals were then fed into our homemade discriminator board based on programmable differential amplifiers (PDAs) LMH6882 and a field programmable gate array (FPGA) Kintex-7 from Xilinx. This board provides dual-threshold differential discriminator and FPGA-based TDC [5] for time measurement based on LED and energy measurement based on time-over-threshold (TOT) method. The threshold of discriminator could be flexibly configured by adjusting threshold setup network. Before the test results were transmitted to PC for further processing, measured events time and energy were fed into a coincidence module implemented in FPGA to reject invalid events. Thus, we constructed a CTR measurement system with abilities of threshold and bias voltage adjustment.

Using the CTR measurement system mentioned above, we performed tests to research on the relationship between CTR and properties such as threshold, SiPM bias voltage and dark count shaping. We measured CTR while bias voltage of SiPM varied from 66.5 V to 68.5 V. For each test, the timing threshold is optimized for better timing performance, and the effect of PZ compensation was tested.

## III. TEST RESULTS

Fig. 3 shows the effect of threshold. This curve is measured under bias voltage of 68V. For LED, lower timing threshold yields better timing resolution. But because of the SiPM

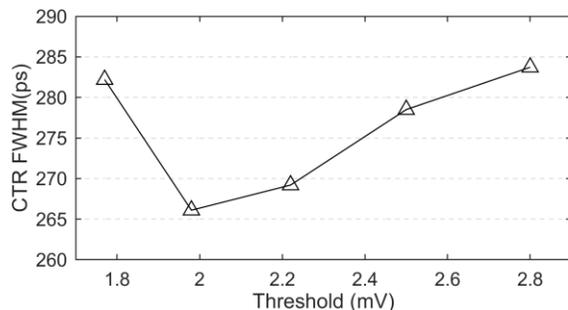

Fig. 3. CTR versus threshold plots. The bias voltage is 68V and the pre-amplifier has PZ compensation network.

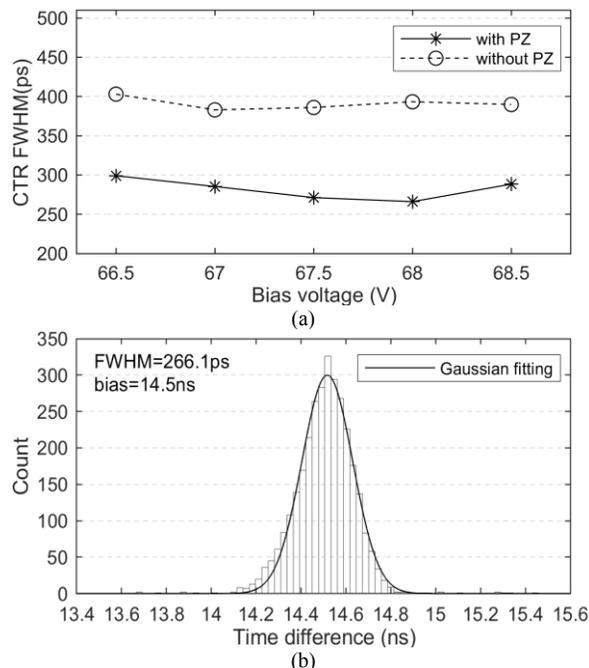

Fig. 4. (a) CTR versus bias voltage plots. The threshold is chosen to get the best CTR under each bias voltage. (b) The histogram of the coincidence time. The pre-amplifier has PZ compensation network.

electronic response and noise, the CTR will be worse if the threshold is too low.

The CTR versus bias voltage curves is shown in Fig. 4(a). As the bias voltage starts from 66.5V, CTR is becoming better due to the increased PDE and gain. After the best point, the negative effect of dark count is comparable with the positive effect, which makes CTR begin to decrease. Considering the PZ structure, two curves show that this structure can improve CTR significantly. Fig. 4(b) shows the histogram and Gaussian fitting of the best CTR result.

## IV. CONCLUSION

Readout circuit of SiPM is very challenging part in TOF-PET detectors. The electronics scheme with the optimized SiPM bias voltage, timing threshold and dark count shaping parameter can significantly improve the timing performance. A CTR of 266ps FWHM has been achieved in our experiment. Considering the other factors limiting CTR, including the size of crystal and operating temperature, our results show excellent performance to the SiPM used in our experiment. This readout circuit will be applied in ASIC design in our next step, and the optimization process is also applicable to other SiPM readout circuits.